\documentclass[11pt,twoside]{article}

\usepackage{asp2010}
\usepackage{graphicx}

\resetcounters 

\begin{document}

\resetcounters

\title{Probing the strong gravity regime with eLISA: Progress on EMRIs}
\author{Carlos F. Sopuerta$^1$ \affil{$ö1$ Institut de Ci\`encies de l'Espai 
(CSIC-IEEC), Campus UAB, Facultat de Ci\`encies, Edifici C5, 2a planta, parells, 
08193 Bellaterra, Spain}}

\begin{abstract} 
The capture of a stellar-mass compact object by a supermassive black hole
and the subsequent inspiral (driven by gravitational radiation emission) 
constitute one of the most important sources of
gravitational waves for space-based observatories like eLISA/NGO.  
In this article we describe their potential as high-precision tools 
that can be used to perform tests of the geometry of black holes
and also of the strong field regime of gravity.  
\end{abstract}

\section{Introduction} 
General Relativity (GR) provides a good description of physical phenomena
that involves gravity for which we have experimental probes, from 
the submillimeter scales to cosmological scales (which require to include
a dark matter and a dark energy component).  However, we have tested 
gravity only in specific regimes (with respect to the strength of the gravitational
field) and for specific length/time scales [a review on experimental tests of 
the validity of GR can be found in~\cite{lrr-2006-3}]. If we look at the strength of
the gravitational field, then we can say that observations to date have 
tested gravity only in the weak and mild gravity regimes, but not in the
strong gravity regime.  To make this statement more precise let us consider
the dimensionless Newtonian potential associated
with a self-gravitating system: $\phi \equiv\phi^{}_{\rm N}/c^{2}$ with
$c$ being the speed of light.
Tests in the weak gravitational regime are those done, for instance, with observations in 
the Solar system (they can measure different precessional effects) and
the typical values of $\phi$ they can reach are\footnote{Here, $G$ denotes the 
gravitational Newton constant and $M^{}_\odot$ the mass of the Sun.}: 
$\phi \sim (G M^{}_\odot)/(c^2\; 1 {\rm AU}) \sim 10^{-8}$\,.
Stronger gravitational fields can be found in pulsars and in particular in binary systems
involving at least one pulsar. In particular, the best evidence
to date of the existence of gravitational radiation comes from observations of 
the well-known Hulse and Taylor binary pulsar (PRS B1913+16)~\cite{Hulse:1974eb,Weisberg:2004hi,Weisberg:2010zz}.
For binary pulsar we have two types of gravitational fields, the self-gravity of the pulsar,
which is very strong and the gravitational attraction of the binary.  The first one is
the self-gravity of a neutron star (NS), and for this we have that $\phi_{1} \sim (G M^{}_{\rm NS})/(c^2\; 
r^{}_{\rm NS})\sim 0.2\,,$ while the second one is (using the Hulse-Taylor pulsar) 
$\phi_{2} \sim (G M^{}_\odot)/(c^2\; r^{\rm periastron}_{\rm Hulse-Taylor}) \sim 10^{-6} \,.$  
By observing the dynamics of the binary system we can test $\phi_{2}$ and also how the
inner structure of the pulsar affects it, i.e. how $\phi_{1}$ may change this dynamics.
It turns out that in General Relativity the inner structure of the individual components
of a binary system only affects the orbital dynamics to the fifth post-Newtonian (PN) order
(\cite{Damour:1987hi}) and hence the effect of $\phi_{1}$ has not been seen but has been used to
put constraints on alternative theories of gravity.  In particular, millisecond pulsar observations
have been used to constraint scalar-tensor theories of gravity, where the internal structure
of the compact objects appears at lower PN order [see, e.g.~\cite{Damour:2007uf,Damour:2007ti,Stairs:2003eg}],
usually due to violations of the strong equivalence principle.  So, we are testing 
gravitational fields with $\phi\sim 10^{-8}$ (Solar system) and $\phi\sim 10^{-6}$ (millisecond pulsars).
In contrast, gravitational wave observations of the merger of compact binaries (NS+NS,NS+BH,
BH+BH) will probe the strong field regime $\phi\sim 1$, which is what we will refer here as
the strong gravitational regime.  

There are different ways to observationally access this regime.  We can try to use
electromagnetic observations, mainly in the high-energy part of the spectrum, 
but there is the drawback that light itself is strongly affected by the
gravitational field of these systems.  Moreover, the electromagnetic emission comes
from matter distributions like accretion disks whose physics is quite complex and 
makes it difficult to produce a precise modeling of the system to be compared with
observations. Nevertheless, tests of the strong field regime have been proposed using
the electromagnetic spectrum (see, e.g.~\cite{Johannsen:2012ta}).
 
Gravitational Waves (GWs) offer an alternative way to access the
strong gravitational regime as they have a weak interaction with matter and hence they carry 
almost uncorrupted information from the systems that produced them.   At present we have several
interferometric ground detectors, like~\cite{LIGO} and~\cite{VIRGO}, 
whose advanced design models will come online around 2014.  They are expected to provide the 
first detections during the present decade, opening the gravitational-wave window in the high-frequency
part of the spectrum, which contains sources like coalescence of compact binaries, supernovae
core collapse, pulsar oscillations, and stochastic backgrounds. 
At the same time, there have been plans for the development of a space-based
interferometric GW observatory, first as an ESA-NASA collaboration
(LISA) and recently as an ESA-only mission, eLISA/NGO (see~\cite{AmaroSeoane:2012je}).
This kind of detectors will operate in the low frequency band (roughly $10^{-4}-1\;$Hz),
which is not accessible from the ground (mainly due to the gravity gradient noise).
The main sources in this part of the spectrum are: coalescence of (super)massive black
holes (MBHs), capture and inspiral of stellar-mass compact objects (SCOs) by a MBH at a
galactic center (also known as {\em Extreme-Mass-Ratio Inspirals} (EMRIs)), 
galactic and extragalactic binaries, and stochastic backgrounds.  
These sources have the potential to provide a wealth of new discoveries that will
impact Astrophysics, Cosmology, and even Fundamental Physics, including constraints
on galaxy formation models, stellar dynamics around galactic nuclei, 
formation and evolution of MBHs, tests of the geometry
of BHs, and even tests of gravity in the strong field regime.  In this write-up we focus
precisely on how to test the strong gravity regime with GW observations
from space of EMRIs.  For a more extended discussion on this see~\cite{Sopuerta:2010zy}.
For recent reviews of the potential of LISA for Fundamental Physics and Cosmology
see:~\cite{Hogan:2007cg,Schutz:2009zz,Schutz:2009tz,Babak:2010ej}.

\section{Extreme-Mass-Ratio Inspirals}
EMRIs are extreme-mass-ratio binaries in the stage where the dynamics
is driven by GW emission.  They are composed of a SCO that inspirals into a MBH located 
in a galactic center.   The masses of interest for the SCO are in the range 
$m^{}_{\star} = 1-10^2\; M^{}_{\odot}$, and for the MBH in the range $M^{}_{\bullet}= 10^5-10^7\; M_{\odot}$.
Then, the mass-ratio for these systems is in the interval: $\mu=m^{}_{\star}/M^{}_{\bullet} \sim 10^{-7} -10^{-3}$.  
As the inspiral proceeds the system losses energy and angular momentum through the emission of 
GWs, producing a secular decay of the orbit and hence a decrease of the orbital periods. 
There are several astrophysical mechanisms that can produce EMRIs (see~\cite{AmaroSeoane:2007aw} for a review
on several aspects of EMRIs).  The most studied mechanism is the gravitational capture of SCOs from a
stellar cusp or core that surrounds the MBH. A number of these SCOs will evolve to a point in which
interactions with other stellar objects are negligible and hence will become EMRIs. 
The EMRIs produced by this mechanism are initially very eccentric, with eccentricities in 
the range  $1-e \sim 10^{-6}-10^{-3}$, but due to GW emission, by the time they enter the band of a space-based
detector it would have been substantially reduced, in the range $e \sim 0.5-0.9$.  

\begin{figure}[!h]
\includegraphics[width=6.8cm]{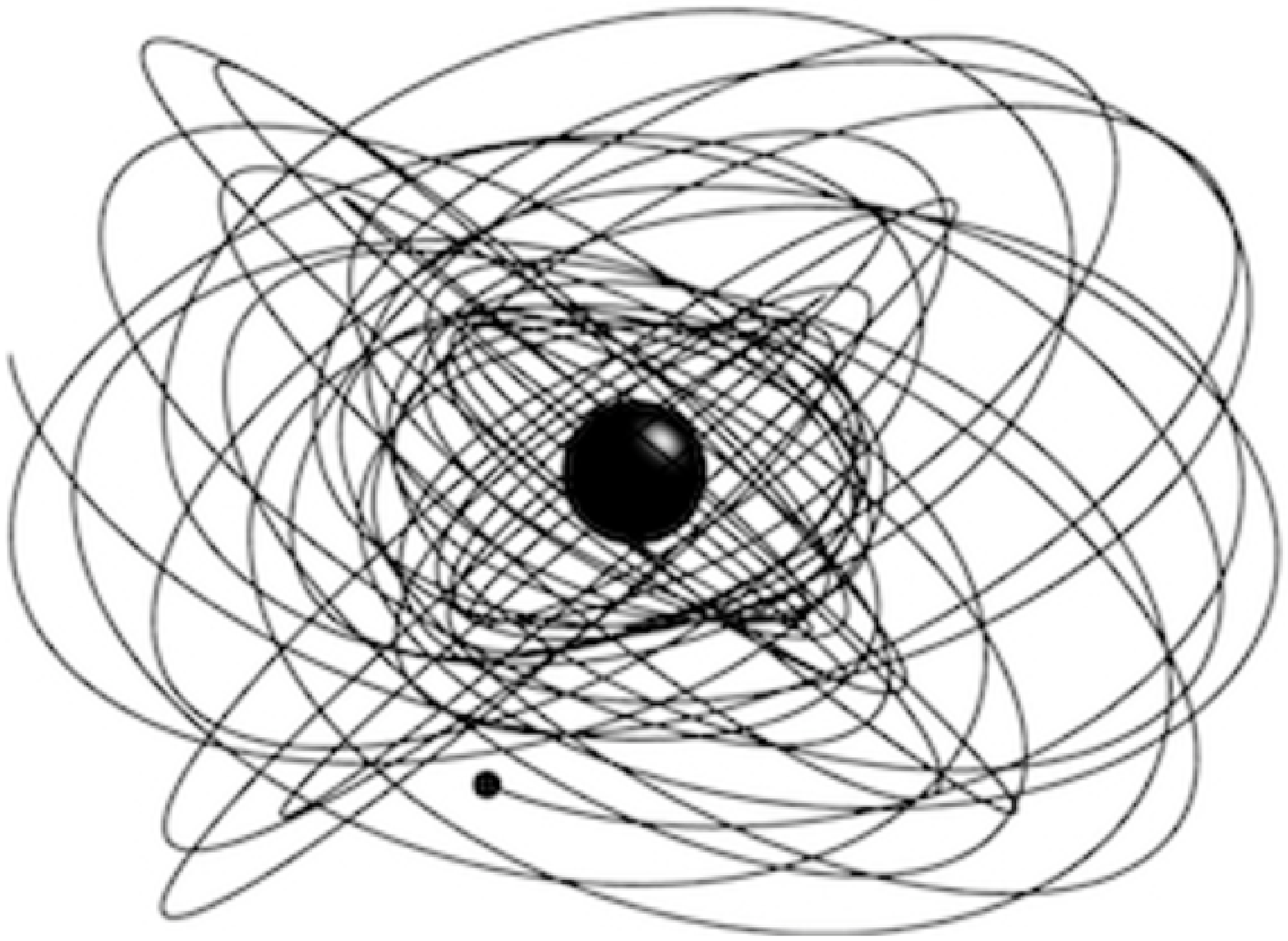}
\includegraphics[width=5.8cm]{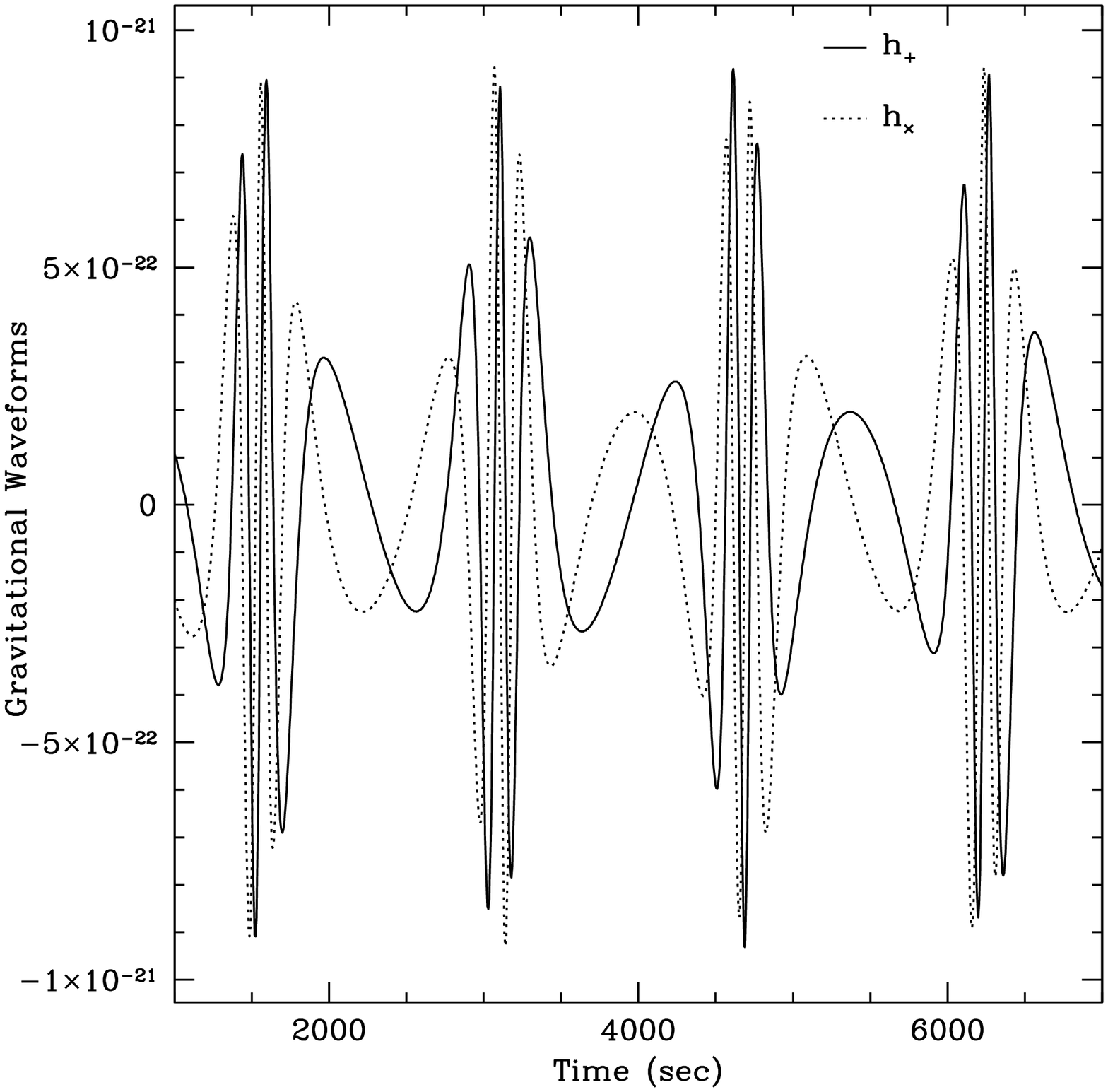}
\caption{\label{fig1}Example of an EMRI trajectory (left) and of an EMRI waveform (right).}
\end{figure}

To understand why EMRIs are useful systems for fundamental physics we have to look at the 
number of cycles they spend in band.  This number scales with the inverse of the mass
ratio, $\mu^{-1}$.  Then, for a detector like LISA, an EMRI can spend more than $10^{5}$ 
GW cycles in band during the last year before plunge (\cite{Finn:2000sy}).  Generic EMRI 
orbits are very complex (see Poisson's contribution on EMRI modeling), 
with high eccentricity and inclined (see Figure~\ref{fig1}), so that they
contain a high number of harmonics contributing to the GW emission (\cite{Barack:2003fp}).
Moreover, many of these cycles can take place very near the MBH horizon, specially for
highly spinning MBHs (see \cite{AmaroSeoane:2012cr}).  All this means that EMRI GWs carry 
a wealth of information about the MBH strong field region.

The next important point is to see how this information about the strong field gravity around
the MBH is encoded in the gravitational waveforms.  The crucial point here is the separation
of scales imposed by the extreme mass ratios involved.   In the case of spatial scales it is
very clear from the ratio of sizes between the MBH and the SCO.  There is also a separation of
time scales due to the dynamics of an EMRI.  The gravitational backreaction, that is, the effect of the
SCO's gravitational field on its own trajectory is the responsible mechanism for the inspiral
of the SCO into the MBH.  Then, we have the orbital time scales (for instance the time to go
from apocenter to pericenter and back) and the inspiral time scale (due to backreaction), whose
ratio is:  $T^{}_{\mbox{\tiny orbital}}/T^{}_{\mbox{\tiny inspiral}}\sim \mu\,$.   This means
that locally in time the inspiral looks like a geodesic trajectory around the MBH.   The {\em constants}
of motion of this geodesic trajectory (which can be taken to be the orbital parameters, $(e,p,\iota)$,
where $p$ is the semilatus rectum and $\iota$ is the inclination angle) change slowly during
the inspiral.  Then, during the long inspiral the trajectory tracks most of the space around the MBH
and close to its horizon.  As a consequence, these long waveforms encode a map of the geometry 
of the MBH (which can be fully parametrized in terms of multiple moments) but also the inspiral
tells us about how the gravitational backreaction mechanism works and hence we have information about 
the details of the theory of gravity.  

To get an idea of the potential of a GW space-based observatory like LISA we can look at estimates
of the parameter estimation studies made for EMRIs.  As we have said, EMRI signals are quite long
($1-2$ yrs or even more) and have significantly high Signal-to-Noise Ratios (SNRs) (SNR $\gtrsim 30$).
Assuming GR and that the no-hair conjecture is true, for a typical EMRI system consisting of an SCO with 
mass $m^{}_{\star} = 10 M^{}_{\odot}$ (a stellar-mass BH)
 inspiralling into a MBH with mass $M^{}_{\bullet} = 10^{6}M^{}_{\odot}$ 
at SNR = $30$, LISA should be able to estimate the main EMRI parameters with the following precision
(see~\cite{Barack:2003fp}; this assumes we see the last year of inspiral):
\begin{equation}
\Delta(\ln M^{}_\bullet)\,,\quad
\Delta\left(\ln\frac{m^{}_{\star}}{M^{}_\bullet}\right)\,, \quad
\Delta\left(\frac{S^{}_\bullet}{M^2_\bullet}\right)~\sim~10^{-4}\,, \label{bcparamestim1}
\end{equation}
where $S^{}_{\bullet}$ denotes the MBH spin (for spinning Kerr BHs: 
$0\leq |S^{}_{\bullet}/M^{2}_{\bullet}|\leq 1$), and
\begin{equation}
\Delta\, e^{}_{o}~\sim~10^{-4}\,,\quad
\Delta\,\Omega^{}_{S}~\sim~10^{-3}\,, \quad
\Delta\,\Omega^{}_{K}~\sim~5\cdot 10^{-2}\,, \label{bcparamestim2}
\end{equation}
where $e^{}_{o}$ is the initial eccentricity, $\Delta\,\Omega^{}_{S}$ is the error in the angular position
of the source in the sky (solid angle), and $\Omega^{}_{K}$ is the ellipse error in the determination
of the MBH spin direction.

Although in this article we are focussing on EMRIs, some of the ideas can, to some extent, also be applied
to Intermediate-Mass-Ratio Inspirals (IMRIs), which instead of a MBH have an intermediate-mass
BH (IMBH) with masses in the range $M^{}_{\bullet}\sim 10^{2-4} M^{}_{\odot}$.  We have not yet conclusive
evidence of the existence of IMBHs but in case they exist, the inspiral of an SCO into an IMBH 
would be detectable by future advanced ground observatories like the Einstein Telescope~\cite{Sathyaprakash:2011bh}.
Moreover, the inspiral of an IMBH into an MBH should be detectable by space-based observatories.  
The modeling of these systems is difficult in the sense that the approximations made
for EMRI modeling are not going to be accurate enough in order to obtain precise IMRI waveform templates
for data analysis purposes.  In particular, the non-linear gravitational dynamics is going to 
play an important role and feedback from Numerical Relativity and post-Newtonian theory is going to be required.

\section{Towards tests of the Strong Gravity Regime with EMRIs}
From the previous discussion it is clear that EMRIs are high precision tools for gravitational
wave astronomy.  The question now is what can be done with them apart from measuring with
precision their physical parameters assuming the standard paradigm, i.e.: GR is the
theory of gravity and the cosmic no-hair conjecture is true and the {\em dark} compact objects at the galactic
nuclei can be described by the Kerr family of solutions~(\cite{Hawking:1973uf}).
The important point is that despite all the observational data available, neither we have experimental 
tests of the strong field gravity nor we have conclusive
evidence that the dark compact objects populating galactic centers are BHs as those described
by the Kerr solutions (even despite the data coming from our own galactic center (\cite{Gillessen:2008qv})).

Then, we can make ourselves several questions.  An obvious one consists in assuming that 
GR is the true theory of gravity and then ask the question of whether the dark objects in
galactic nuclei have a geometry compatible with the Kerr solution. We will call this the
{\em Kerr hypothesis} and the main problem is to figure out how to test it in practice.  
The solution must contain a method to discriminate between the Kerr geometry and other possible 
geometries for collapsed objects, in particular for the case of EMRI signals.
As an starting point we are going to assume that the geometry of the dark compact objects is 
stationary, asymptotically flat, and axisymmetric.  Then, within GR, it can be shown
that the geometry can be fully characterized by two sets of numbers, the multipole moments,
$M^{}_{\ell}$ and $J^{}_{\ell}$ ($\ell = 0, 1,\ldots$).  The first set, $\{M^{}_{\ell}\}$,
are the mass moments, which also exist in Newtonian gravity.  The second set, $\{J^{}_{\ell}\}$,
are the current moments, which do not exist in Newtonian gravity.  In GR the current moments
are a consequence of the fact that not only the mass density gravitates but also the 
momentum density.  The Kerr multipole moments satisfy the following 
elegant relations:
\begin{equation}
M^{}_{\ell} + i\, J^{}_{\ell} = M^{}_{\bullet}\, \left(i\, a^{}_{\bullet}\right)^{\ell} \,,\label{kerrmoments}
\end{equation}
and as expected, they are fully determined by the BH mass $M^{}_{\bullet}$ and the BH intrinsic 
angular momentum (spin) $S^{}_{\bullet}$ ($a^{}_{\bullet} = S^{}_{\bullet}/M^{}_{\bullet}$).   
That is, only $M^{}_{0}$ and $J^{}_{1} = M^{}_{\bullet}a^{}_{\bullet}$ are independent moments, the rest is 
a combination of them.  For instance,
the next mass and current multipoles can be related to these two by the following simple relations:
$- M^{}_{\bullet}a^{2}_{\bullet} = M^{}_{2} = - J^{2}_{1}/M^{}_{0}\,$ and
$- M^{}_{\bullet}a^{3}_{\bullet} = J^{}_{3} = M^{}_{2}J^{}_{1}/M^{}_{0}\,.$
This discussion about the multipole moments can be used to rewrite the Kerr hypothesis in a way more 
suitable for tests using GW EMRI observations: {\em Kerr Hypothesis}: 
The exterior gravitational field of the dark, compact and very massive 
objects sitting at the galactic centers can be well described by the vacuum, stationary, and axisymmetric 
solutions of the General Theory of Relativity whose multipole moments 
$\{M^{}_{\ell},J^{}_{\ell}\}^{}_{\ell=0,\ldots,\infty}$ satisfy the Kerr relations of Eq.~(\ref{kerrmoments}).

What remains is to see how to do this in practice with GW EMRI observations.  A first method
to experimentally test the strong-field predictions of general relativity was proposed by~\cite{Poisson:1996ep} 
using a simplified model where the inspiral is a sequence of circular orbits in the equatorial plane of the MBH.
Later, Ryan,  in a series of 
papers (\cite{Ryan:1995wh,Ryan:1995xi,Ryan:1997hg}), studied how to
test experimentally the Kerr hypothesis using the multipole moments.
In a first study~\cite{Ryan:1995wh} studied to what extend the GW emission of EMRIs 
depends on the values of the different multipole moments of the 
central massive object. To that end, Ryan assumed nearly circular and nearly equatorial orbits 
and that the inspiral takes place in a slowly and adiabatic manner, avoiding analyzing
the GW emission in detail.  In particular, Ryan showed that the number of cycles that the 
dominant GW components spend in a logarithmic frequency interval, $\delta N(f) \equiv
f^{2}/(df/dt)$, which contains equivalent information to the GW phase, 
contains full information of the whole set of multipole moments. 
Ryan also showed how to extract at least the
first three moments, ($M^{}_{0}$, $J^{}_{1}$, $M^{}_{2}$), which is enough to perform a partial test of
the Kerr hypothesis.  

In a second study~(\cite{Ryan:1997hg}) the goal was to estimate
the accuracy of LIGO and LISA in determining the multipole moments from the GWs 
emitted by an IMRI (LIGO) or an EMRI (LISA). The basic idea
is to use waveform template models that include a certain number of multipole moments 
in the set of source parameters.  It is well known that standard EMRI GWs can be 
described in terms of $14$ parameters, hence the waveforms used by Ryan are time
series $h(t;{\theta}^{I})$, with $\theta^{I} = \theta^{I}_{\rm GR}$ ($I=1,\ldots,14$)
begin the standard EMRI parameters and $\theta^{I} = \theta^{I}_{M}$ ($I=15,\ldots,14+N^{}_{mm}$),
are the extra parameters containing $N^{}_{mm}$ multipole moments (different from the first two, $M^{}_{0}$ and
$J^{}_{1}$).  The conclusions of Ryan's estimations indicate that a space-based detector
like LISA can be able to make several tests of the Kerr hypothesis.  In the particular case 
of an SCO with $m^{}_{\star} = 10 M^{}_{\odot}$ inspiralling into a central massive
object with $M=10^{5}M^{}_{\odot}$, during 2 years before plunge and with an 
SNR of $10$, Ryan found that LISA could
measure $M$ with a fractional error of $\Delta M/M\sim 10^{-3}$, $m^{}_{\star}$ with fractional error 
of $\Delta m^{}_{\star}/m^{}_{\star} \sim 10^{-3}$,
and for the spin Ryan finds $\Delta(S/M^{2})\sim 10^{-3}\,.$ Finally, for the mass quadrupole moment 
the accuracy found is $\Delta(M^{}_{2}/M^{3}) \sim 0.5$.  Adding many multipole moments in the analysis
degrades the precision significantly.  

More recently,~\cite{Barack:2006pq} did a similar analysis with some improvements
with respect to Ryan's work.   First of all, the waveform model
is essentially the same used to produce the estimations of Eqs.~(\ref{bcparamestim1}) 
and~(\ref{bcparamestim2}).  In this model, the EMRI system follows, at any instant in time, a Newtonian 
orbit emitting lowest-order quadrupolar GWs. Nevertheless, the model contains post-Newtonian modifications to 
secularly evolve the parameters of the orbit in such a way that the model contains all the features 
that we expect from generic EMRIs.  The LISA model used by Barack \& Cutler is also more complete than Ryan's one, 
including the main features of the LISA constellation motion and a better noise model that includes the confusion
noise from galactic and extragalactic binaries.  However, Barack \& Cutler only include the effects of
the mass quadrupole $M^{}_{2}$.  The parameter estimation analysis of~\cite{Barack:2006pq}
predicts that the mass quadrupole could be measure with errors in the range 
[for an EMRI system as in Eqs.~(\ref{bcparamestim1}) and~(\ref{bcparamestim2}) but with SNR $=100$]:
\begin{equation}
\Delta (M^{}_{2}/M^{3}_{0})~ \sim~ 10^{-4} - 10^{-2}\,,
\end{equation}
which is a much better error estimate than Ryan's one, mainly due to the full complexity of the EMRI
dynamics encoded in the waveform model of~\cite{Barack:2003fp}.

The program initiated by Ryan is far from being complete.  One question that arises is to what extend
the multipole moment parametrization of the central object should be carried forward to the parametrization
of the EMRI GW signals.  First of all, the multipole moments are quantities defined at spatial infinity,
far away from the strong field region, which complicates the way in which the generated GWs depend on them.
Secondly, the expansion of the gravitational field in terms of multipole moments is such that higher
multipoles will contribute less and less to the expansion as long as we evaluate the expansion far away
enough from the central compact object.  However, we have mentioned that EMRIs can expend a large fraction
of cycles very near the last stable orbit, which in turn can be quite close to the horizon for spinning
MBHs and prograde orbits.  This means that for these situations a high number of multipoles are 
going to be relevant for the
description since most powers of $M/r$ are going to be of the same order near pericenter, 
where the field and the GW emission are stronger.  Then, the multipole moments as waveform parameters 
are going to exhibit correlations
and the error estimations for them are not going to be good enough for making precise tests.  
In summary, it is possible that a space-based GW detector will measure a few multipole moments with high
precision, allowing for partial tests of the Kerr hypothesis, but including too many multipoles in the
waveform model may be counterproductive for these purposes.    

Other ideas to test the strong field region near the dark compact objects in galactic
nuclei have been proposed. For instance,~\cite{Collins:2004na} 
initiated a program to construct exact solutions within GR that are almost BHs (they are
stationary, axisymmetric, asymptotically flat and vacuum, and have been named  
{\em bumpy} BHs) but such that some multipole moments have the {\em wrong} value.  An 
interesting feature of these geometries is that they are valid in the strong field 
region.   These bumpy BHs may be
used for performing {\em null} experiments, i.e. by comparing their properties 
with measurements of astrophysical sources.
The drawback of the initial solutions was that the changes introduced are not 
smooth and present some pathological strong-field structure. These problems were fixed
by~\cite{Vigeland:2009pr} who also introduced angular momentum (bumpy Kerr BHs).  
Construction of other bumpy BH geometries has continued and has been extended to other
theories of gravity (\cite{Vigeland:2011ji}).

There are other studies that consider geometries for the central compact object 
alternative to Kerr. \cite{Kesden:2004qx} investigated the possibility of distinguishing between a central
MBH and a boson star from signatures in the waveforms produced by the inspiral of an SCO. 
The main idea is that for a MBH the waveform will end at the plunge whereas for the boson star 
it will continue until the SCO reaches its center. 
\cite{Barausse:2006vt} constructed models that add a self-gravitating and homogeneous compact torus 
to the MBH with comparable mass and spin. It was found that for most cases the emission from
these  systems are indistinguishable 
from pure-Kerr waveforms, indicating a possible confusion problem.  
Moreover,~\cite{Barausse:2007dy} find that in general the dissipative effect of the hydrodynamic 
drag exerted by the torus on the SCO is much 
smaller than the one due to radiation reaction.  Other exact solutions, like the Manko-Novikov solutions
(\cite{Manko:1992mn}) have been used (\cite{Gair:2007kr,Apostolatos:2009vu,LukesGerakopoulos:2010rc}),
where the orbital motion can exhibit chaotic behavior and this can produce 
observable signatures in the EMRI waveforms.

Up to know we have discussed several ways of testing deviations from the standard EMRI
paradigm within GR.  Mainly by studying changes in the geometry of the central compact
object (one could also study changes in the SCO structure, like adding spin~(\cite{Huerta:2011kt}),
although these are in general less important). 
In what follows, we discuss how to extend these studies by allowing 
for theories that include a description of the gravitational interaction different from
the one of GR.  In this sense, it is important to mention that the landscape of theories
of gravity is very rich.  However, it is not always possible to extract what are the 
predictions of a given theory for a system like an EMRI, either because of the complexity
of the theory or because we lack consistent formulations of the problem.
Nevertheless, there are several theories of gravity different from GR where EMRIs have
been studied.  For instance in the well-known class of scalar-tensor theories of gravity,
where it has been suggested that EMRIs may exhibit {\em floating} orbits (\cite{Yunes:2011aa}), 
a very different behavior with respect to GR.  To illustrate the potential of EMRI GW observations, 
in what follows we describe a couple of examples where EMRIs have 
been studied in two different theories of gravity and the predictions that have been made 
for LISA observations.  

The first example is about EMRIs is Dynamical Chern-Simons Modified Gravity (DCSMG; see
\cite{Alexander:2009tp} for a review). This is a modification of GR introduced first 
by~\cite{jackiw:2003:cmo} that consists in
adding to the GR action a gravitational parity-violating term, the Pontryagin invariant, 
which is a generalization
of the well-known three-dimensional Chern-Simons (CS) term.   Since in four dimensions the 
Pontryagin term is a topological invariant and would not contribute to the field equations,
in DCSMG it appears coupled to a scalar field.  To avoid the restrictive dynamics that comes
out of the original model (see~\cite{Yunes:2007ss}), in DCSMG we also add the action for this
scalar field.  This theory can be seen as a low-energy limit of string theory or even 
of loop quantum gravity, but it can be seen just as a gravitational correction in the spirit 
of effective field theories.

There are several interesting points about this theory with regard to the description
of EMRIs. First, spinning MBHs in this theory are no longer described by the 
Kerr metric (although non-spinning MBHs are still described by the Schwarzschild metric).  
There are deviations from Kerr that have been found using a slow-rotation and 
a small-coupling (these deviations are controlled by a single parameter, $\xi$, that turns out 
to be a combination of universal coupling constants) approximation (\cite{Yunes:2009hc,Yagi:2012ya}). 
These corrections affect the  multipolar structure of the MBH but, as it happens with 
the Kerr metric, they still remain completely determined by only two numbers, the mass $M^{}_{\bullet}$ 
and spin parameter $a^{}_{\bullet}$.  Therefore, the main spirit of the no-hair conjecture, as far 
as we know, remains valid. For the lowest order correction (order $\xi\cdot a^{}_{\bullet}$),  
the Kerr relations for the multipole moments satisfy different relations for $\ell \geq 4$, 
which involve the CS parameter $\xi$.  However, when we add the next correction (order
$\xi\cdot a^{2}_{\bullet}$, these relations are modified for lower $\ell$ (\cite{Yagi:2012ya}).
In what follows, we will describe only results found using the lowest-order correction,
whose details can be found in~\cite{Canizares:2012is} and in~\cite{Sopuerta:2009iy}. 
One important point is that in this theory the effective energy-momentum tensor of the GWs 
is formally as in GR, with contributions to the radiation reaction mechanism from the CS scalar field.
Moreover, in DCSMG distant observers will notice the same GW polarizations as in GR
(\cite{Sopuerta:2009iy}).

The local dynamics of EMRIs in DCSMG is very similar to the one in GR in the sense
that the modified MBH metric has the same symmetries (at order $\xi\cdot a^{}_{\bullet}$)
as the Kerr metric.  Therefore, locally in time, the waveforms are dominated by multiples of three fundamental
frequencies and one could think that these can lead to a confusion problem between GR and
DCSMG EMRI waveforms.  What breaks the {\em degeneracy} are the radiation reaction effects.
Recently, a parameter estimation study of EMRIs has been carried out 
(see \cite{Canizares:2012is} and also Canizares' contribution to this volume~(\cite{Canizares:2012he})).
In the case where DCSMG is assumed to be the {\em true} theory of gravity and for an
EMRI with $M^{}_{\bullet}=10^{6}M^{}_{\odot}\,,$ $a^{}_{\bullet}/M^{}_{\bullet}=0.25\,,$
$e^{}_{0}=0.25\,,$ and $\xi\,a^{}_{\bullet}/M^{5}_{\bullet}=5\cdot 10^{-2}\,,$ 
it was found that the expected errors are in the range: 
$\Delta\log M^{}_{\bullet} \sim 5\cdot10^{-3}\,,$ 
$\Delta a^{}_{\bullet}\sim 5\cdot10^{-6}\,,$ 
$\Delta e^{}_{0}\sim 3\cdot10^{-7}\,,$ and
$\Delta\log(\xi\cdot a^{}_{\bullet})\sim 4\cdot10^{-2}\,.$
In contrast, if we assume that GR is the {\em true} theory of gravity and that measurements
are compatible with a vanishing $\xi$, we should be able to establish, using LISA EMRI measurements, 
the following bound on this CS parameter
(\cite{Canizares:2012is}): $\xi^{1/4} < 1.4\cdot 10^{4}{\rm km}\,,$
which is almost four orders of magnitude better than the bound imposed by Solar System 
experiments.

A completely different study of EMRIs has been done in the context of the higher-dimensional
braneworld scenarios of~\cite{Randall:1999vf}.  In these scenarios, standard model fields
live on a 3+1 brane moving in a five-dimensional spacetime where the direction out of the brane
is not non-compact.  They were proposed as an attempt to solve the hierarchy problem of the
physical {\em fundamental} interactions. 
They are also a playground for holography (and the anti-de Sitter/Conformal
Field Theory (AdS-CFT) correspondence) since it 
has been argued that the BH solutions localized on the brane, found by solving the classical 
bulk equations correspond to quantum-corrected black holes, rather than classical ones.
As a consequence, it was conjectured that static black holes can not exist for radius much 
greater than the Anti-de Sitter (AdS) length of the extra dimension, $L$.
The presence of a large number of CFT degrees of freedom accelerates the decay of a black hole via 
Hawking radiation determining its evolution. It has been found (\cite{Emparan:2002jp}) that
$d{m}^{}_{\bullet}/dt \approx -2.8\times 10^{-3} (M^{}_{\odot}/m^{}_{\bullet})^2 
(L/1mm)^2 M^{}_{\odot}\, {\rm yr}^{-1}\,$.
Therefore, EMRIs where the SCO is a stellar-mass BH can be used to constraint this theory.
\cite{McWilliams:2009ym} found that LISA could constraint $L$ to be below $5$ microns.
However, static solutions for black holes have been found recently (\cite{Figueras:2011gd})
invalidating the hypothesis on which these constraints are based.  
These hypothesis are based on applying free field theory intuition to the CFT on the brane, 
which is strongly coupled, and this may explain the apparent contradiction.
In any case, the important lesson from this example is that EMRI GW observations have 
a tremendous potential to test fundamental physics scenarios
linked with high-energy physics (for more on this see~\cite{Cardoso:2012qm}).

\section{Conclusion} 
The main goal of this article has been to convey the idea that observations of EMRIs by
future space-based observatories like eLISA will provide a unique opportunity to make
great discoveries in the area of Fundamental Physics, specially for learning about the
geometry of the presumed MBHs at the galactic centers and about the strong field regime of
gravity that so far has been experimentally unexplored.

\section*{Acknowledgments}
CFS acknowledges support from the Ram\'on y Cajal Programme of the Spanish Ministry of Education 
and Science, contract 2009-SGR-935 of AGAUR,
and contracts FIS2008-06078-C03-03, AYA-2010-15709, and FIS2011-30145-C03-03 of MICCIN. 
We acknowledge the computational resources provided by the BSC-CNS
(AECT-2011-3-0007) and CESGA (contracts CESGA-ICTS-200 and CESGA-ICTS-221).


\begin{thebibliography}{}
\expandafter\ifx\csname natexlab\endcsname\relax\def\natexlab#1{#1}\fi
\expandafter\ifx\csname url\endcsname\relax
  \def\url#1{\texttt{#1}}\fi
\expandafter\ifx\csname urlprefix\endcsname\relax\def\urlprefix{URL }\fi
\providecommand{\eprint}[2][]{\url{#2}}

\bibitem[{Alexander \& Yunes(2009)}]{Alexander:2009tp}
Alexander, S., \& Yunes, N. 2009, Phys. Rep., 480, 1. \eprint{0907.2562}

\bibitem[{Amaro-Seoane et~al.(2012{\natexlab{a}})Amaro-Seoane, Sopuerta, \&
  Freitag}]{AmaroSeoane:2012cr}
Amaro-Seoane, P., Sopuerta, C.~F., \& Freitag, M.~D. 2012{\natexlab{a}}.
  \eprint{1205.4713}

\bibitem[{Amaro-Seoane et~al.(2007)}]{AmaroSeoane:2007aw}
Amaro-Seoane, P., et~al. 2007, Class. Quant. Grav., 24, R113.
  \eprint{astro-ph/0703495}

\bibitem[{Amaro-Seoane et~al.(2012{\natexlab{b}})}]{AmaroSeoane:2012je}
--- 2012{\natexlab{b}}, Class. Quant. Grav., 29, 124016. \eprint{1202.0839}

\bibitem[{Apostolatos et~al.(2009)Apostolatos, Lukes-Gerakopoulos, \&
  Contopoulos}]{Apostolatos:2009vu}
Apostolatos, T.~A., Lukes-Gerakopoulos, G., \& Contopoulos, G. 2009, Phys. Rev.
  Lett., 103, 111101. \eprint{0906.0093}

\bibitem[{Babak et~al.(2011)Babak, Gair, Petiteau, \& Sesana}]{Babak:2010ej}
Babak, S., Gair, J.~R., Petiteau, A., \& Sesana, A. 2011, Class. Quant. Grav.,
  28, 114001. \eprint{1011.2062}

\bibitem[{Barack \& Cutler(2004)}]{Barack:2003fp}
Barack, L., \& Cutler, C. 2004, Phys. Rev., D69, 082005. \eprint{gr-qc/0310125}

\bibitem[{Barack \& Cutler(2007)}]{Barack:2006pq}
--- 2007, Phys. Rev., D75, 042003. \eprint{gr-qc/0612029}

\bibitem[{Barausse \& Rezzolla(2008)}]{Barausse:2007dy}
Barausse, E., \& Rezzolla, L. 2008, Phys. Rev., D77, 104027. \eprint{0711.4558}

\bibitem[{Barausse et~al.(2007)Barausse, Rezzolla, Petroff, \&
  Ansorg}]{Barausse:2006vt}
Barausse, E., Rezzolla, L., Petroff, D., \& Ansorg, M. 2007, Phys. Rev., D75,
  064026. \eprint{gr-qc/0612123}

\bibitem[{Canizares et~al.(2012{\natexlab{a}})Canizares, Gair, \&
  Sopuerta}]{Canizares:2012he}
Canizares, P., Gair, J.~R., \& Sopuerta, C.~F. 2012{\natexlab{a}}.
  \eprint{1209.2534}

\bibitem[{Canizares et~al.(2012{\natexlab{b}})Canizares, Gair, \&
  Sopuerta}]{Canizares:2012is}
--- 2012{\natexlab{b}}, Phys. Rev., D86, 044010. \eprint{1205.1253}

\bibitem[{Cardoso et~al.(2012)}]{Cardoso:2012qm}
Cardoso, V., et~al. 2012. \eprint{1201.5118}

\bibitem[{Collins \& Hughes(2004)}]{Collins:2004na}
Collins, N.~A., \& Hughes, S.~A. 2004, Phys. Rev., D69, 124022.
  \eprint{gr-qc/0402063}

\bibitem[{Damour(1987)}]{Damour:1987hi}
Damour, T. 1987, in {Three hundred years of gravitation}, edited by {Hawking,
  S. W. and Israel, W.} ({Cambridge}: {Cambridge University Press})

\bibitem[{Damour(2007{\natexlab{a}})}]{Damour:2007uf}
--- 2007{\natexlab{a}}, in {Physics of Relativistic Objects in Compact
  Binaries: from Birth to Coalescence}, edited by {Colpi, M. and Casella, P.
  and Gorini V. and Moschella, U. and Possenti, A.} ({Berlin}: {Springer}).
  \eprint{0704.0749}

\bibitem[{Damour(2007{\natexlab{b}})}]{Damour:2007ti}
--- 2007{\natexlab{b}}, in {Revised Edition of: Neutron Stars, Black Holes and
  Binary X-Ray Sources}, edited by H.~Gursky, \& R.~Ruffini ({Dordrecht}: {D.
  Reidel Pub. Co.}). \eprint{0705.3109}

\bibitem[{Emparan et~al.(2003)Emparan, Garcia-Bellido, \&
  Kaloper}]{Emparan:2002jp}
Emparan, R., Garcia-Bellido, J., \& Kaloper, N. 2003, JHEP, 0301, 079.
  \eprint{hep-th/0212132}

\bibitem[{Figueras \& Wiseman(2011)}]{Figueras:2011gd}
Figueras, P., \& Wiseman, T. 2011, Phys. Rev. Lett., 107, 081101.
  \eprint{1105.2558}

\bibitem[{Finn \& Thorne(2000)}]{Finn:2000sy}
Finn, L.~S., \& Thorne, K.~S. 2000, Phys. Rev., D62, 124021.
  \eprint{gr-qc/0007074}

\bibitem[{Gair et~al.(2008)Gair, Li, \& Mandel}]{Gair:2007kr}
Gair, J.~R., Li, C., \& Mandel, I. 2008, Phys. Rev., D77, 024035.
  \eprint{0708.0628}

\bibitem[{Gillessen et~al.(2009)}]{Gillessen:2008qv}
Gillessen, S., et~al. 2009, Astrophys.J., 692, 1075. \eprint{0810.4674}

\bibitem[{Hawking \& Ellis(1973)}]{Hawking:1973uf}
Hawking, S.~W., \& Ellis, G. F.~R. 1973, {The Large Scale Structure of
  Space-Time} (Cambridge: Cambridge University Press)

\bibitem[{Hogan(2007)}]{Hogan:2007cg}
Hogan, C.~J. 2007. \eprint{0709.0608}

\bibitem[{Huerta \& Gair(2011)}]{Huerta:2011kt}
Huerta, E., \& Gair, J.~R. 2011, Phys.Rev., D84, 064023. \eprint{1105.3567}

\bibitem[{Hulse \& Taylor(1975)}]{Hulse:1974eb}
Hulse, R.~A., \& Taylor, J.~H. 1975, The Astrophysical Journal, 195, L51

\bibitem[{Jackiw \& Pi(2003)}]{jackiw:2003:cmo}
Jackiw, R., \& Pi, S.~Y. 2003, Phys. Rev., D68, 104012. \eprint{gr-qc/0308071}

\bibitem[{Johannsen(2012)}]{Johannsen:2012ta}
Johannsen, T. 2012. \eprint{1209.3024}

\bibitem[{Kesden et~al.(2005)Kesden, Gair, \& Kamionkowski}]{Kesden:2004qx}
Kesden, M., Gair, J., \& Kamionkowski, M. 2005, Phys. Rev., D71, 044015.
  \eprint{astro-ph/0411478}

\bibitem[{{LIGO}()}]{LIGO}
{LIGO}{.~Laser Interferometer GW Observatory. URL: {\tt www.ligo.caltech.edu}}

\bibitem[{Lukes-Gerakopoulos et~al.(2010)Lukes-Gerakopoulos, Apostolatos, \&
  Contopoulos}]{LukesGerakopoulos:2010rc}
Lukes-Gerakopoulos, G., Apostolatos, T.~A., \& Contopoulos, G. 2010, Phys.
  Rev., D81, 124005. \eprint{1003.3120}

\bibitem[{{Manko} \& {Novikov}(1992)}]{Manko:1992mn}
{Manko}, V.~S., \& {Novikov}, I.~D. 1992, Class. Quant. Grav., 9, 2477

\bibitem[{McWilliams(2010)}]{McWilliams:2009ym}
McWilliams, S.~T. 2010, Phys. Rev. Lett., 104, 141601. \eprint{0912.4744}

\bibitem[{Poisson(1996)}]{Poisson:1996ep}
Poisson, E. 1996, Phys.Rev., D54, 5939. \eprint{gr-qc/9606024}

\bibitem[{Randall \& Sundrum(1999)}]{Randall:1999vf}
Randall, L., \& Sundrum, R. 1999, Phys. Rev. Lett., 83, 4690.
  \eprint{hep-th/9906064}

\bibitem[{Ryan(1995)}]{Ryan:1995wh}
Ryan, F.~D. 1995, Phys. Rev., D52, 5707

\bibitem[{Ryan(1996)}]{Ryan:1995xi}
--- 1996, Phys. Rev., D53, 3064. \eprint{gr-qc/9511062}

\bibitem[{Ryan(1997)}]{Ryan:1997hg}
--- 1997, Phys. Rev., D56, 1845

\bibitem[{Sathyaprakash et~al.(2012)}]{Sathyaprakash:2011bh}
Sathyaprakash, B., et~al. 2012, Class. Quant. Grav., 29, 124013.
  \eprint{1108.1423}

\bibitem[{Schutz(2009)}]{Schutz:2009zz}
Schutz, B.~F. 2009, Class. Quant. Grav., 26, 094020

\bibitem[{Schutz et~al.(2009)Schutz, Centrella, Cutler, \&
  Hughes}]{Schutz:2009tz}
Schutz, B.~F., Centrella, J., Cutler, C., \& Hughes, S.~A. 2009.
  \eprint{0903.0100}

\bibitem[{{Sopuerta}(2010)}]{Sopuerta:2010zy}
{Sopuerta}, C.~F. 2010, GW Notes, 4, 3. \eprint{1009.1402}

\bibitem[{Sopuerta \& Yunes(2009)}]{Sopuerta:2009iy}
Sopuerta, C.~F., \& Yunes, N. 2009, Phys. Rev., D80, 064006. \eprint{0904.4501}

\bibitem[{Stairs(2003)}]{Stairs:2003eg}
Stairs, I.~H. 2003, Living Reviews in Relativity, 6. \eprint{astro-ph/0307536}

\bibitem[{Vigeland et~al.(2011)Vigeland, Yunes, \& Stein}]{Vigeland:2011ji}
Vigeland, S., Yunes, N., \& Stein, L. 2011, Phys. Rev., D83, 104027.
  \eprint{1102.3706}

\bibitem[{Vigeland \& Hughes(2010)}]{Vigeland:2009pr}
Vigeland, S.~J., \& Hughes, S.~A. 2010, Phys. Rev., D81, 024030.
  \eprint{0911.1756}

\bibitem[{{VIRGO}()}]{VIRGO}
{VIRGO}{.~URL: {\tt www.virgo.infn.it}}

\bibitem[{Weisberg et~al.(2010)Weisberg, Nice, \& Taylor}]{Weisberg:2010zz}
Weisberg, J., Nice, D., \& Taylor, J. 2010, Astrophys. J., 722, 1030.
  \eprint{1011.0718}

\bibitem[{Weisberg \& Taylor(2005)}]{Weisberg:2004hi}
Weisberg, J.~M., \& Taylor, J.~H. 2005, ASP Conf. Ser., 328, 25.
  \eprint{astro-ph/0407149}

\bibitem[{Will(2006)}]{lrr-2006-3}
Will, C.~M. 2006, Living Reviews in Relativity, 9. \eprint{gr-qc/0510072}

\bibitem[{Yagi et~al.(2012)Yagi, Yunes, \& Tanaka}]{Yagi:2012ya}
Yagi, K., Yunes, N., \& Tanaka, T. 2012, Phys. Rev., D86, 044037.
  \eprint{1206.6130}

\bibitem[{Yunes et~al.(2012)Yunes, Pani, \& Cardoso}]{Yunes:2011aa}
Yunes, N., Pani, P., \& Cardoso, V. 2012, Phys. Rev., D85, 102003.
  \eprint{1112.3351}

\bibitem[{Yunes \& Pretorius(2009)}]{Yunes:2009hc}
Yunes, N., \& Pretorius, F. 2009, Phys. Rev., D79, 084043. \eprint{0902.4669}

\bibitem[{Yunes \& Sopuerta(2008)}]{Yunes:2007ss}
Yunes, N., \& Sopuerta, C.~F. 2008, Phys. Rev., D77, 064007. \eprint{0712.1028}

\end{thebibliography}

 \end{document}